\documentclass[twocolumn,showpacs,preprintnumbers,amsmath,amssymb]{revtex4}

\usepackage{xspace}
\usepackage{amsfonts}
\usepackage{graphicx}
\usepackage{url}
\usepackage[dvipdfm,%
              colorlinks,%
          linkcolor=blue,%
        citecolor = blue,%
              hyperindex,%
        plainpages=false,%
         urlcolor = blue,%
       pdfstartview=FitH]{hyperref}

\usepackage{amsfonts}

\begin{document}


\title{\textsf{Enhancement of Carrier Mobility in Semiconductor Nanostructures by Dielectric Engineering}}

\author{Debdeep Jena \& Aniruddha Konar}
\affiliation{Department of Electrical Engineering and Department of Physics \\
            University of Notre Dame, IN 46556}
\date{\today}

\begin{abstract}
We propose a technique for achieving large improvements in carrier
mobilities in 2- and 1-dimensional semiconductor nanostructures by
modifying their dielectric environments.  We show that by coating
the nanostructures with high-$\kappa$ dielectrics, scattering from
Coulombic impurities can be strongly damped.  Though screening is
also weakened, the damping of Coulombic scattering is much larger,
and the resulting improvement in mobilities of carriers can be as
much as an order of magnitude for thin 2D semiconductor membranes,
and more for semiconductor nanowires.
\end{abstract}

\pacs{73.50.-h} \keywords{Semiconductor Nanostructures, Mobility,
Dielectric Mismatch, Scattering}

\maketitle



The dielectric mismatch between a nanoscale semiconductor material
(relative dielectric constant $\epsilon_{s}$) and the surrounding
environment ($\epsilon_{e}$) can result in a number of peculiarities
not present in their bulk and layered forms (such as epitaxially
grown quantum wells and superlattices).  For example, recently it
has been shown that in metallic carbon nanotubes, there exist
so-called `tubular image states' due to the strong attraction
between external charges with their image states in the
tube\cite{grangerPRL02, zamkovPRL04, segalPRL05}. Furthermore, the
dielectric mismatch effect has been shown to be responsible for
polarization-sensitive photocurrents in nanowires
\cite{wangNature01, shabaevNL04}. Keldysh had predicted from theoretical 
considerations that the Coulombic interaction potential between electrons 
and holes in nanoscale thin freestanding ($\epsilon_{e}$=1) semiconductor 
films \cite{keldyshJETP79}, and in freestanding nanowires\cite{keldyshPSS97}
is strongly modified by the dielectric environment, resulting in a
large enhancement in the excitonic binding energy.  His prediction was 
recently experimentally confirmed\cite{muljarovPRB00, dneprovskiiJETP02}. 
Goldoni et al. proposed enhancement of excitonic binding energy in 
dielectric-semiconductor heterostructures by separate dielectric confinement of electric
fields, and quantum-confinement of carriers\cite{goldoniPRL98}. Due
to the leakage of electric field lines originating from charges in
the semiconductor nanostructure into the surrounding, a large degree
of {\em tunability} of the optical properties of semiconductor
nanostructures is possible by designing the dielectric environment.

\begin{figure}
\begin{center}
\leavevmode 
\includegraphics[width=3in]{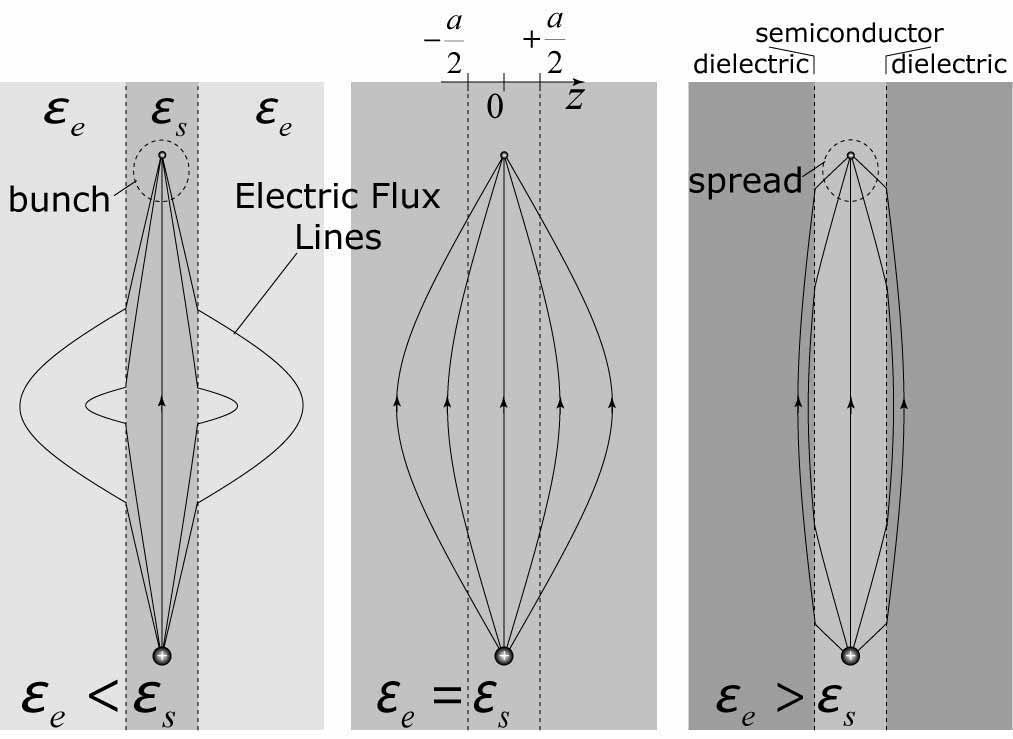}
\caption{Electric flux lines originating from a fixed ionized
impurity and terminating on a mobile electron, and the effect of the
dielectric environment.  The flux lines bunch closer inside the
semiconductor layer if $\epsilon_{e} < \epsilon_{s}$, and spread
farther apart if $\epsilon_{e} > \epsilon_{s}$, thus enhancing
Coulomb interaction in the former case and damping it in the latter.} 
\label{fig:Fig1}
\end{center}
\end{figure}

The introduction of the concept of modulation doping by St\"{o}rmer,
Dingle, et al in 1978 for epitaxially grown semiconductor
heterostructures resulted in a major advance in achieving very high
mobility 2-dimensional electron gases (2DEGs)\cite{dingleAPL78}.
Such 2DEGs have resulted in various fundamental discoveries in
transport physics, and have also found many applications in
high-speed transistor design.  Over the last decade, major strides
have been made in the bottom-up synthesis of various nanostructures,
such as 0-dimensional (0D) nanocrystal quantum
dots\cite{talapinScience05}, 1D nanowires\cite{xiangNature06}, and
2D nanoscale semiconductor sheets (for example,
graphene\cite{ohtaScience06, obradovicAPL06, bergerScience06,
novoselovNature05, zhangPRL05}, nanoribbons\cite{khangScience06},
etc), which do not require the stringent lattice-matching requirements placed
by epitaxy. In this work, we propose a novel technique for improving the
electron mobility in such bottom-up 1- and 2-D semiconductor nanostructures by
showing that the dielectric environment has a profound effect on
electron transport.  Semiconductor nanowires\cite{xiangNature06} and
nanosheets\cite{obradovicAPL06} are currently being investigated for
possible usage in technologically important electronic devices such
as transistors.  Carrier mobility is an important parameter for such
devices since it determines the power dissipation and switching
speed.  Proper design of the dielectric environment can result in
drastic improvements in carrier mobilities in such structures.  The
availability of high-quality insulating materials with a very large
range of relative dielectric constants ($1 < \epsilon_{e} < 300$)
implies that the proposed technique of mobility improvement can be
implemented with current technology.


We consider a thin 2D semiconductor membrane of nanoscale thickness
in this work.  To find the effect of the dielectric environment on
electron mobility in the membrane, we first investigate its effect
on Coulombic scattering of mobile charges from ionized impurity
atoms located inside the membrane.  A point charge $e$ is located at
$(0,0,z_{0})$ inside the semiconducting membrane of thickness $a$
(see Fig \ref{fig:Fig1}).


The dielectric mismatch and electric field penetration into the
surrounding results in a potential inside the membrane that can be
found by a Green's function solution of Poisson equation, or by the
method of image charges. For the system here, an infinite series of
point image charges form at points $ z_{n} = n a + (-1)^{n}z_{0},$
where $n=\pm 1, \pm 2, \ldots$.  The $n^{th}$ point charge has a
magnitude $e\gamma^{|n|}$, where $\gamma =
(\epsilon_{s}-\epsilon_{e})/(\epsilon_{s}+\epsilon_{e})$ is a
measure of the dielectric mismatch. The unscreened potential due to
the Coulomb scatterer experienced by a mobile electron at a point
$({\vec{\rho}},z)$ inside the membrane is then given
by\cite{kumagaiPRB89}

\vspace{-2ex}
\begin{equation}
V_{unsc}^{Coul}({\bf \rho},z) = \sum_{n=-\infty}^{\infty}
\frac{e\gamma^{|n|}}{4\pi \epsilon_{0} \epsilon_{s} \sqrt{|{\vec{
\rho}}|^{2} + [z-z_{n}]^{2}} }. \label{totpotential}
\end{equation}
\vspace{-2ex}

The image charges are all of the same sign if $\epsilon_{e} <
\epsilon_{s}$, and the effective potential seen by electrons in the
membrane is {\em larger} than if there were no dielectric mismatch.
The sign of image charges oscillate between positive and negative
when $\epsilon_{e} > \epsilon_{s}$, and the total effective Coulomb
potential seen  is therefore {\em smaller} than if $\epsilon_{e} =
\epsilon_{s}$ (shown schematically in Figure \ref{fig:Fig1}).

Assuming that the surrounding dielectric provides a large energy
barrier for confining electrons in the semiconductor membrane, the
envelope function of mobile electrons occupying the $n_{z}^{th}$
subband is given by $\Psi_{n_{z},{\bf k}}({\vec{\rho}}, z) =
\phi_{n_{z}}(z) \chi_{\bf k} ({\vec{ \rho}})$, where
$\phi_{n_{z}}(z) = \sqrt{2/a} \cos (\pi n_{z} z/a)$, $\chi_{\bf k}(
\vec{\rho} ) = \exp({i {\bf k} \cdot {\vec {\rho}}})/\sqrt{S}$,
$\hbar{\bf k}$ being the in-plane quasi-momentum, and $S$ the area
of the membrane. The matrix element for scattering from state
$|n_{z}, {\bf k_{i}}\rangle \rightarrow |m_{z}, {\bf k_{f}}\rangle$
due to the unscreened Coulomb potential in Eq. \ref{totpotential} is
given by an infinite sum of the Fourier transforms of the potentials
due to the point charge and all it's image charges
$\widetilde{V}_{unsc}^{Coul}(q) = \sum_{n= -\infty}^{\infty}
\widetilde{v}_{n}(q)$.  Here $\widetilde{v}_{n}(q)$ is the matrix
term for the $n^{th}$ image charge, and is calculated to be

\vspace{-2ex}
\begin{equation}
\widetilde{v}_{n}(q,z) = \frac{e^{2}}{2 \epsilon_{0}\epsilon_{s}q}
\cdot e^{-q|z-z_{n}|} \cdot \gamma^{|n|} \cdot F_{n_{z},m_{z}}(aq),
\end{equation}
\vspace{-2ex}

where $q=|{\bf k_{i} - k_{f}}|$, and $F_{n_{z},m_{z}}(aq)$ is a
form-factor arising from the quasi-2D nature of the electron gas.
The form factor can be exactly evaluated for the envelope function
used. $F_{n_{z},m_{z}}(aq) \rightarrow 1$ both in the long
wavelength limit, as well as for very thin membranes ($a \rightarrow
0$).  For thin membranes, the approximation $z_{0} = 0$ is
justified; under this approximation, the total contribution of all
point charges can be exactly summed. We calculate the total
unscreened potential seen by the mobile electron to be

\vspace{-2ex}
\begin{equation}
\widetilde{V}_{unsc}^{Coul}(q) = \frac{e^{2}}{2 \epsilon_{0}
\epsilon_{s} q } \cdot F_{ n_{z}, m_{z} }(aq) \cdot [\frac{e^{qa} +
\gamma }{e^{qa} - \gamma}] , \label{totCoulMatElement}
\end{equation}
\vspace{-2ex}

where the factor in rectangular brackets arises entirely due to the
dielectric mismatch.  In the long-wavelength limit, this factor
approaches $\epsilon_{s}/\epsilon_{e}$, indicating that {\em the
unscreened Coulomb interaction within the film is dominated by the
dielectric constant of the environment
($\widetilde{V}_{unsc}^{Coul}(q) \sim e^{2}/2 \epsilon_{0}
\epsilon_{e} q$) and not the semiconducting membrane}.



The other scattering mechanism that affects the mobility of carriers
in very thin films is due to surface roughness (SR).  Due to
statistical fluctuations of atomic migration during the growth
process, the thickness of such thin membranes can fluctuate randomly
over monolayer thicknesses over the surface.  Prange \& Nee have
shown that the roughness in the film thickness can be modeled by a
function $\Delta a ({\bf \rho})$, which has a correlation $\langle
\Delta a ({\bf \rho}) \Delta a ({\bf \rho + \rho^{\prime}})
\rangle_{\rho} =\Delta^{2} \exp (-|\rho -
\rho^{\prime}|^{2}/\Lambda^{2})$, where $\Delta, \Lambda$ are the
film thickness variation, and the in-plane correlation length
between the fluctuations respectively\cite{prangePhysRev68,
sakakiAPL87}. With such a model, the unscreened matrix element for
SR intrasubband scattering for electrons within the $m_{z}^{th}$
subband in the 2DEG is given by

\vspace{-2ex}
\begin{equation}
\widetilde{V}_{unsc}^{SR}(q) = \frac{ \pi^{5/2} \hbar^{2} }{
m^{\star} } \cdot \frac{ \Delta \Lambda m_{z}^{2} }{ a^{3} } \cdot
e^{- \frac{ (q\Lambda)^{2} }{ 8 } }, \label{totSRMatElement}
\end{equation}
\vspace{-2ex}

where $\hbar$ is the reduced Planck's constant, and $m^{\star}$ is
the electron effective mass.



The carrier mobility in the nanoscale membranes will be determined
by the combined effect of Coulomb and SR scattering.  For reasonably
high impurity doping which is necessary for large intrinsic
conductivity of the membrane, phonon scattering is weaker, and the
reason for neglecting it is explained later.  A careful look at the
effect of the dielectric environment on the screening of the
scattering potentials derived in Eqs. \ref{totCoulMatElement} \&
\ref{totSRMatElement} is necessary for the evaluation of the total
scattering rates. Screening by 2-dimensional electrons is captured
in the $T \rightarrow 0$ limit by the Lindhard
function\cite{sternPRL67}

\vspace{-2ex}
\begin{equation}
\epsilon_{2d}(q,\omega) = 1 + \frac{e^{2}}{2 \epsilon_{0}
\epsilon_{s} q } \Pi (q,\omega) \Phi(q,a),
\end{equation}
\vspace{-2ex}

where $\Pi$ is the polarization function and $\Phi(q,a)$ is a factor
that depends on the subband index of screening carriers,
$\epsilon_{s}$, $\epsilon_{e}$, membrane thickness $a$, and $q$.  In
the static Lindhard limit, the polarization function reduces to the
2D density of states $\Pi (q,\omega) \rightarrow m^{\star}/\pi
\hbar^{2}$, and the dielectric mismatch factor can be written as
$\Phi = \Phi_{1} + \Phi_{2}$ (see ref.\cite{glavinPRB02}).
$\Phi_{1}(q, a)$ depends only on the scattering wavevector $q$ and
the semiconductor film thickness $a$, whereas
$\Phi_{2}(q,a,\epsilon_{s},\epsilon_{e})$ also depends on the
dielectric mismatch.  We have evaluated the exact analytical
expressions for $\Phi_{1}$ and $\Phi_{2}$ for the nanoscale
membrane.  The expressions are long, and will be presented in a more
detailed paper; here, we mention some important features of the
environment-modified dielectric screening function.  For the rest of
this work, we consider the electric quantum limit, i.e., for
intra-subband scattering within the 1st subband ($n_{z} = m_{z} =
1$).

\begin{figure}
\begin{center}
\leavevmode 
\includegraphics[width=3in]{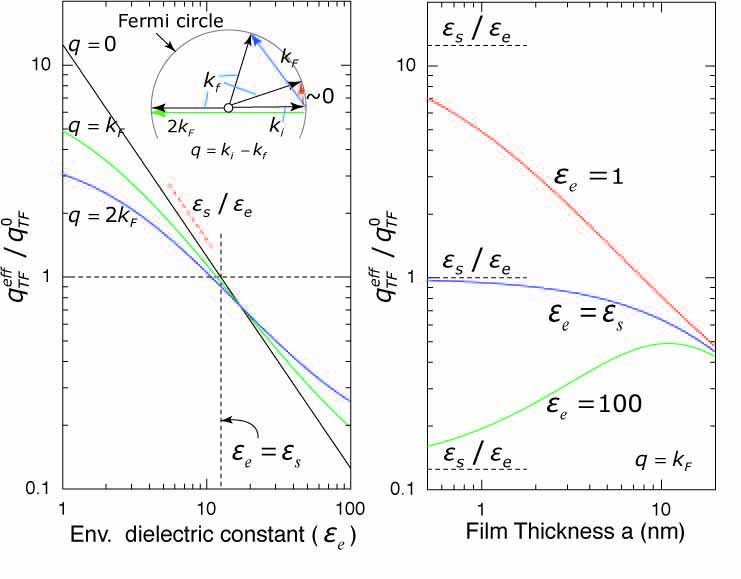}
\caption{The effect of dielectric mismatch on the Thomas-Fermi
screening wavevector. (a) shows that screening is stronger than the
case without dielectric mismatch for $\epsilon_{e} < \epsilon_{s}$
for small angle scattering ($q \sim 0$), $q = k_{F}$, and
backscattering $q = 2k_{F}$ (for $a=1$nm). When $\epsilon_{e}
> \epsilon_{s}$, screening is weaker than the case without
dielectric mismatch. (b) shows that the effect of dielectric
mismatch on screening is substantial if the thickness of membrane is
small.} 
\label{fig:Fig2_Screening}
\end{center}
\end{figure}

In the long wavelength limit, $\Phi_{1} \rightarrow 1$, and
$\Phi_{2} \rightarrow \epsilon_{s}/\epsilon_{e}-1$.  The screening
function for a general $(q,a)$ is written as $
\epsilon_{2d}^{eff}(q) = 1 + q_{TF}^{eff} / q $ in analogy to the 2D
screening function in the absence of dielectric mismatch ($
\epsilon_{2d}(q) = 1 + q_{TF}^{0} / q $).  Here
$q_{TF}^{0}$($q_{TF}^{eff}$) is Thomas-Fermi screening wavevector
for the 2DEG without(with) dielectric mismatch.  The ratio
$q_{TF}^{eff}/q_{TF}^{0}$ that captures the effect of the dielectric
mismatch on screening is plotted in Figs \ref{fig:Fig2_Screening} a
\& b for a GaAs thin film ($m^{\star} = 0.067 m_{0}$, $\epsilon_{s}
= 12.5$). Finite-temperature effect ($T=300$K) has been considered
using Maldague's technique (see ref \cite{sakakiAPL87}).  As can be
seen, screening is {\em enhanced if $\epsilon_{e} < \epsilon_{s}$},
and {\em reduced if $\epsilon_{e}
> \epsilon_{s}$} due to the dielectric mismatch.  Evidently,
screening opposes the change in Coulomb interaction between charges
due to the dielectric environment.  Furthermore, the ratio
approaches $\epsilon_{s}/\epsilon_{e}$ at the limit of very small
film thickness, implying that {\em screening in a very thin 2D film
is mediated entirely by the dielectric environment}.



The dielectric-mediated screening changes each scattering potential
to $\widetilde{V}_{scr}^{i}(q) =
\widetilde{V}_{unsc}^{i}(q)/\epsilon_{2d}^{eff}(q)$, and the
momentum scattering rate for the $i^{th}$ scattering mechanism is
given by

\vspace{-2ex}
\begin{equation}
\frac{1}{\tau_{i}(E_{k})} = \frac{2 \pi}{\hbar} \int
\frac{d^{2}k^{\prime}}{(2 \pi)^{2}} |\widetilde{V}_{scr}^{i}|^{2} (
1 - \cos \theta ) \delta ( E_{k} - E_{k^{\prime}}),
\end{equation}
\vspace{-2ex}

where $\cos \theta = {\bf k_{i} \cdot k_{f}}/{\bf |k_{i}| \cdot
|k_{f}|}$, $k = |{\bf k_{i}}|$, $k^{\prime}=|{\bf k_{f}}|$, and the
scattering rate is evaluated over the final density of states. A
parabolic dispersion with effective mass $m^{\star}$ is used. Since
both Coulombic and SR scattering are elastic, Matheissen's rule can
be applied to evaluate the total scattering rate ($1/\langle \tau
\rangle = 1/\tau_{Coul} + 1/\tau_{SR}$), and the Drude-mobility is
then given by $\mu = e \langle \tau \rangle / m^{\star}$.

\begin{figure}
\begin{center}
\leavevmode 
\includegraphics[width=3in]{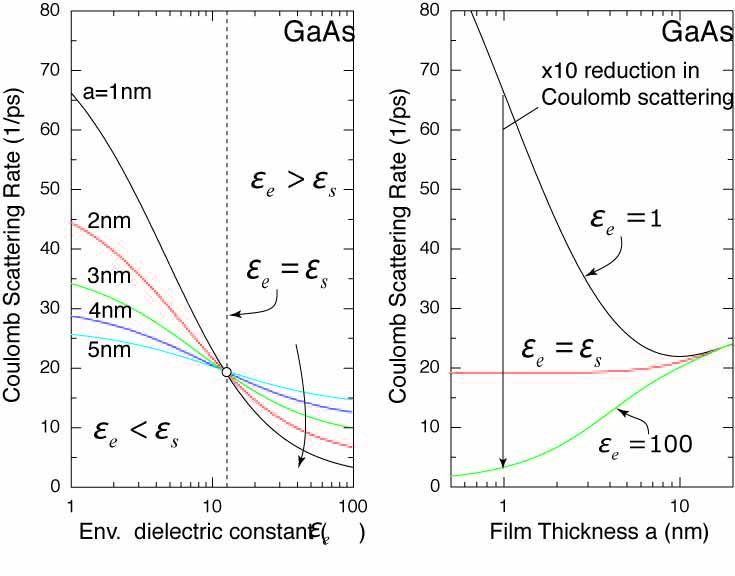}
\caption{(a) -
The variation of Coulombic scattering rate for different
semiconductor layer thicknesses for a range of dielectric
environments.  (b) - Effect of dielectric mismatch on Coulombic
scattering as a function of the semiconductor membrane thickness.} 
\label{fig:Fig3_CoulScat}
\end{center}
\end{figure}

We first investigate the effect of the dielectric environment on
Coulomb scattering.  The scattering rate for a 2DEG of density
$n_{2d} = 10^{12}/$cm$^{2}$ due to randomly distributed Coulombic
point scatterers of the same density ($n_{imp} = n_{2d}$) in a
nanoscale GaAs membrane is shown  Fig \ref{fig:Fig3_CoulScat}. The
carrier density chosen ensures the electric quantum limit for the
range of membrane thicknesses considered here.  Fig
\ref{fig:Fig3_CoulScat}(a) shows the effect on the scattering rate
as the dielectric constant of the environment is changed.  For very
thin layers ($a=1$nm), the scattering rate exhibits a very strong
dependence on $\epsilon_{e}$, reducing $\sim 20$ times from 66/ps to
3/ps as the environmental dielectric constant changes from
$\epsilon_{e}=1$ (semiconductor film in air) to $\epsilon_{e}=100$
(film coated by a high-$\kappa$ dielectric).  As the film thickness
increases, the effect of the dielectric mismatch on Coulombic
scattering diminishes, and this feature is shown explicitly in
Fig.\ref{fig:Fig3_CoulScat}(b).  As the film thickness approaches $a
= 20$nm, Coulombic scattering becomes bulk-like, and the dielectric
mismatch has no effect beyond this thickness.

Hence Coulombic scattering in membranes below a critical thickness
(given in the long-wavelength limit by $a_{cr} \sim a_{B}^{\star}
\epsilon_{e}/(\epsilon_{e}+\epsilon_{s})$, $a_{B}^{\star}$ being the
hydrogenic Bohr-radius in the bulk semiconductor) can be strongly
suppressed by coating the semiconductor layer with insulating
dielectric barriers with high $\epsilon_{e} (> \epsilon_{s})$.  On
the other hand, a free-standing ($\epsilon_{e} = 1$) doped
semiconductor membrane will suffer from enhanced Coulombic
scattering.  For a GaAs nanoscale film, the strength of Coulombic
scattering can be reduced by more than a factor of 20 for $a=1$nm.
Similar results can be obtained for membranes made of other
semiconducting materials.

\begin{figure}
\begin{center}
\leavevmode 
\includegraphics[width=3in]{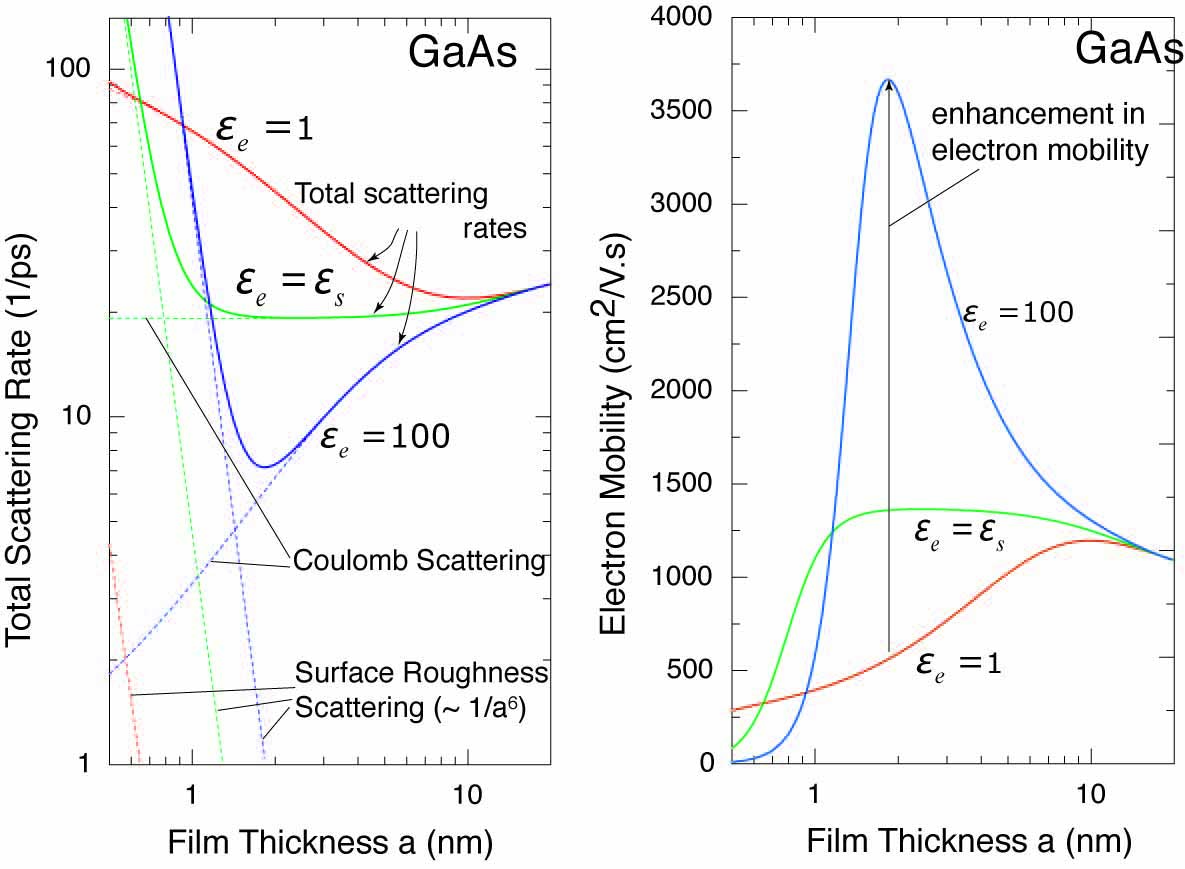}
\caption{(a)
- Total scattering rate due to combined surface roughness and
Coulombic impurity scattering, and (b) - the dependence of the total
mobility for three dielectric environments on the semiconductor
membrane thickness.  A large increase in mobility is expected for
thin semiconductor layers coated by high-$\kappa$ dielectrics.} 
\label{fig:Fig4_Mobility}
\end{center}
\end{figure}

In Fig \ref{fig:Fig4_Mobility}(a), we plot the total scattering rate
by combining the effects of surface roughness and Coulombic
scattering for three different dielectric environments.  The surface
roughness parameters used are 1 monolayer thickness variations
$\Delta = 0.283$nm, with an in-plane correlation length $\Lambda =
40$nm (these values are typical for GaAs quantum wells (see Ref.
\cite{sakakiAPL87}).  The 2D electron gas density and the impurity
density are both set to $10^{12}$/cm$^{2}$.  Under such conditions,
impurity and SR scattering is found to be much stronger than polar
optical phonon scattering (LO phonon scattering rate for GaAs is
$\sim 2$/ps at room temperature), justifying its exclusion in the
analysis. Since SR scattering has a $a^{-6}$ dependence on the
membrane thickness, it dominates for $a<1$nm for GaAs.  The strong
dependence of SR scattering on the dielectric environment is
entirely due to the dielectric mismatch effect on screening, since
the unscreened SR scattering matrix element (Eq.
\ref{totSRMatElement}) is not affected by the dielectric mismatch.

When the two processes are combined, the total scattering rate is
lower compared to the bulk limit if $\epsilon_{e} > \epsilon_{s}$,
and the thickness of the membrane is $a_{min} < a < a_{cr}$, where
$a_{min}$ is set by the surface roughness. For atomically flat
surfaces (for example graphene), $a_{min}=0$nm. The corresponding
electron mobilities for the GaAs membrane are plotted in Fig
\ref{fig:Fig4_Mobility}(b). For $a>1$nm, the mobility can be
improved by coating the film by high-$\kappa$ dielectrics; an enhancement
as large as by a factor of 7 is possible by choosing ($\epsilon_{e}
= 100$) for the GaAs membrane considered.  The mobility is enhanced
in general for ($\epsilon_{e}
> \epsilon_{s}$) due to dielectric-mediated lowering of Coulomb
scattering within the thickness window, the lower limit of which is
set by the surface roughness, and the upper limit by the effective
Bohr radius.  Thus, large enhancements in carrier mobility are
achievable in this fashion for membranes made from a wide range of
semiconductors.


Relaxing the infinite barrier approximation leads to an increase of the electron effective mass, 
which can reduce the mobility.  This correction is numerically evaluated for typical barrier 
heights ($\geq$1 eV), and found to be small enough to be neglected.  The dielectric constant of a dielectric which has only an {\em electronic} contribution to the polarization scales inversely with the (direct) bandgap\cite{zimanBOOK}.  Therefore, to 
provide the confining potential for electrons assumed in this work, the high-$\kappa$ dielectrics 
will necessarily be of the type in which the {\em ionic} contribution is substantially larger than the electronic contribution.  
Choice of such dielectrics (primarily oxides) implies that the polaronic dressing of the electron 
effective mass might offset the enhancement in mobility predicted.  The polaronic increase in the effective mass is evaluated to be less than 10\% for a wide range of ionic dielectrics, implying that the dielectric-mismatch induced giant enhancement in mobility persists.  


In conclusion, doped nanostructures suffer from enhanced Coulombic impurity scattering if they are freestanding.  The electron mobility in such structures can be improved drastically by coating them with high-$\kappa$ dielectrics. 
The large sensitivity of carrier mobilities in such membranes to the dielectric environment can be exploited for sensor applications.  Finally, we note that the effect of dielectric mismatch on carrier scattering and mobility in ultrathin semiconductor nanowires is expected to be even stronger than in the nanoscale membranes considered here due to much stronger field penetration into the environment and weaker screening in 1D transport than in 2D. Therefore, the improvement of carrier mobilities for semiconductor nanowires by dielectric engineering will be higher than what is found here for 2D membranes.  This is an attractive route for improving the performance of nanostructured electronic devices.

\end{document}